\documentclass[letterpaper]{sfchem}
\usepackage{graphicx}

\begin{document}

\title{Future Mm/Submm Instrumentation and Science Opportunities: Example of Deuterated Molecules}

\author{Thomas G.\ Phillips\inst{1} \and Charlotte Vastel\inst{1}} 
  \institute{California Institute of Technology, 320-47, Pasadena, CA  91125, U.S.A.} 
\authorrunning{Phillips, Vastel}
\titlerunning{Future mm and submm instrumentation}

\maketitle 

\begin{abstract}

During the next decade a tremendous advance will take place in instrumentation for spectroscopy of 
the interstellar medium. Major new facilities (ALMA, SOFIA, APEX, LMT, Herschel and others) will be 
constructed and commissioned, so that the science opportunities, in the field of astrochemistry, will 
increase by a huge factor. This will be enhanced by the new receivers with greater bandwidth and 
sensitivity. The new opportunities will be in the area of astrochemistry of distant objects, 
through greater sensitivity, or new spectral ranges due to the platforms above the Earth's atmosphere.\\
Various aspects of new spectral ranges are discussed, with emphasis on H$_2$O lines, features previously 
hidden under H$_2$O or O$_2$ lines, light hydrides and particularly on deuterium in molecules. Recently, 
multiply deuterated species have been detected, e.g. ND$_3$, in cold dense regions of the interstellar 
medium. It is argued here that it is possible that so much deuterium could be trapped, by the 
fractionation process, into heavy molecules such as ND$_3$, etc..., and species such as H$_2$D$^+$ 
and possibly D$_2$H$^+$, that D and HD might be depleted. This would be the mechanism for the large 
dispersion of [D]/[H] values found in the interstellar medium. Light molecules (hydrides and deuterides) 
generally have large fundamental rotation frequencies, often lying in the HIFI bands. 
The deuterides are a specially suitable case, because the species exist mainly 
in cold dense regions, where the molecules are in the ground states and THz observations will best be 
carried out by absorption spectroscopy against background dust continuum sources such as Sgr B2 and W49N. 


\end{abstract}

\section{Introduction}
  
\begin{figure}[ht]
\resizebox{\hsize}{!}{\includegraphics{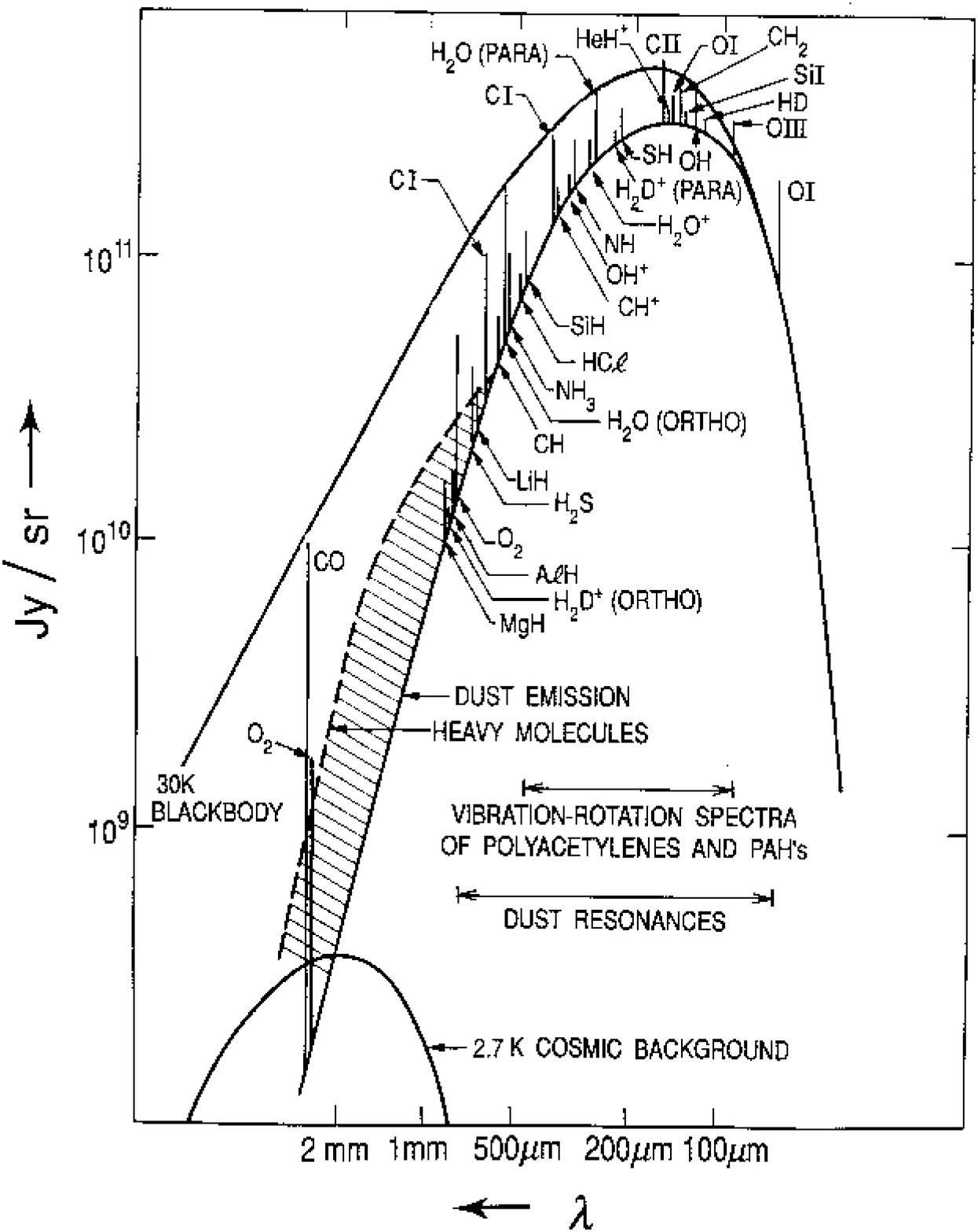}}
\caption{Illustration of the principal spectral lines expected at submillimeter wavelengths, 
shown superposed on the dust continuum emission of a 30 K interstellar cloud.
\label{phillips_keene}}
\end{figure}

A great deal is already known about the nature of the gas and dust in the interstellar medium, 
obtained from a wide variety of telescopes and facilities. Most of these have 
been ground-based, but some airborne, balloon-borne, or space missions. In particular, high 
resolution, heterodyne (R $\ge$ 10$^6$) spectrometers have provided information on the 
molecular and (surprisingly) atomic species in the various types of interstellar clouds. 
It has indicated the physical nature of the gas, mostly from the velocity structure and tested 
the chemical models, e.g., the ion-molecule reaction scheme, which was of course developed to 
explain
the observed molecular content of the clouds. Figure \ref{phillips_keene} (\cite{phillips92}) 
shows a submillimeter and far-infrared spectrum of an interstellar cloud, as it might be 
observed from a space platform. Features include
the line-forest of heavy molecules at the longer wavelengths, light hydrides and deuterides 
at shorter wavelengths,
important atoms and ions, such as C$^0$, C$^+$, O$^0$... and dust continuum and resonances. 
Also large molecules are possibly to be seen via their bending modes.\\

\begin{figure}[ht]
\resizebox{\hsize}{!}{\includegraphics{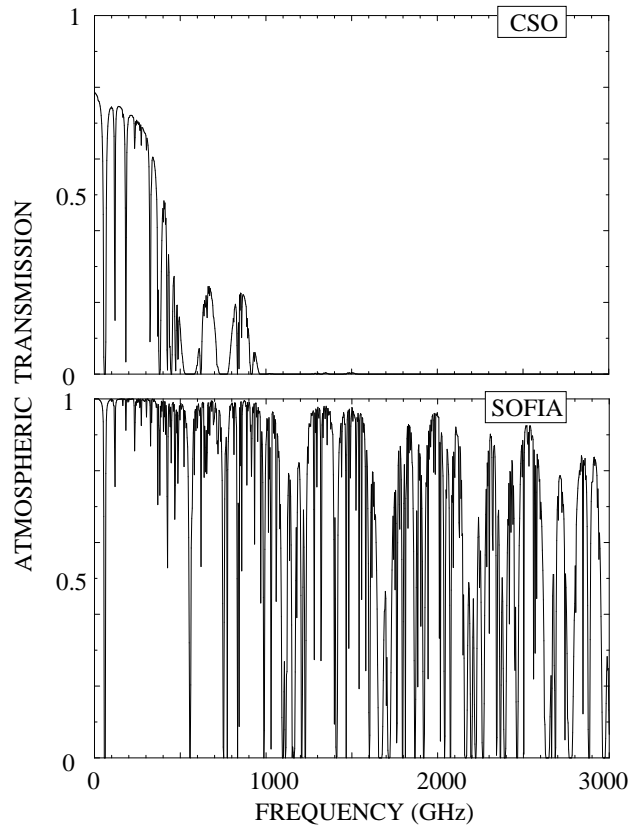}}
\caption{Comparison of the atmospheric transmissions between the sky visible at CSO (with a standard 
precipitable water column of 1 mm) and the sky that will be visible for SOFIA (with a standard 
precipitable water column of 
10 $\mu$m), both at an elevation of 45$^{o}$.
\label{atm}}
\end{figure}

Compared to other branches of astronomy, the submillimeter has been slow to develop, in part due to
the low photon energy (compared to optical) and the small number of arriving photons (compared to
the radio), making it difficult to construct detection equipment, but also due to the poor overall
transmission of the Earth's atmosphere. As is seen in Figure \ref{atm} (top), even from a high
mountain site such as Mauna Kea, Hawaii, the transmission is only partial up to about 900 GHz (330
$\mu$m) and very little above that. Of course, from airborne altitudes, i.e., 12 km for SOFIA,
transmission is much better (Figure \ref{atm}, bottom) but is still badly interrupted by H$_2$O and
O$_2$ lines. There is a clearly indicated need for space-based telescopes.\\

A third, and possibly dominant reason for the underdeveloped nature of the field is the lack of 
major instruments. Because the millimeter/submillimeter observers are developing their field later, 
in time, compared to the optical, 
radio and X-ray communities, we have had to wait our turn for major funding. The wait is
almost over and the full potential of this exciting field is about to be realized in the coming decade!

\section{New Instruments}

We are concerned with the new opportunities for interstellar chemistry, which include investigation of
molecules and atoms in the gas phase, dust grains and surface chemistry and the interactions
between the gas and dust phases, including accretion of molecules onto grains, and the opposite
process, where the accretion material is returned to the gas phase. Here, we discuss this only from
the point of view of gas spectroscopy.\\
Table \ref{instruments} lists many of the major new telescopes and interferometers which will be
available within the next decade (hopefully).

\begin{table*}
\begin{center}
\caption{New facilities for interstellar chemistry.}\vspace{1em}
\renewcommand{\arraystretch}{1.2}
\begin{tabular}[h]{cccc}
\hline
Name      &   Aperture       &   Platform                 &   Available   \\
\hline
SOFIA     &   2.5 m          &  Airborne (747)            &  2005\\
Herschel  &   3.5 m          &    Space (L2)              &  2007\\
ALMA      & 64 $\times$ 12 m & Atacama (Chile)            &  2010\\
CARMA     & 6 $\times$ 10 m  & White Mts (California, USA)&  2005\\
SMA       & 8 $\times$ 6 m   & Mauna Kea (Hawaii, USA)&  2003\\
e-SMA     &(8 $\times$ 6 m) + 15 m + 10 m & Mauna Kea (Hawaii, USA)&  2005\\
LMT       &   50 m           & Sierra Negra (Mexico)      &  2004\\
APEX      &   12 m           & Atacama (Chile)            & 2005\\
South Pole&   8 m           & South Pole                 & 2006\\
GBT       &  100 m           & West Virginia (USA)        &  2002\\
\hline \\
\end{tabular}
\label{instruments}
\end{center}
\end{table*}
 
Compared to what we have today, this will be a staggering new capability and the field will lead
astronomy in that it will discover and define the new objects and new concepts to be
investigated. One might worry that there will not be enough trained people in the field to fully
utilize such a cornucopia of facilities, so university-based facilities should expand their
efforts in this field now!\\
ALMA represents the biggest step forward, providing an unprecedented spectroscopic capability, 
throughout the atmospheric windows, for nearby and distant objects. This huge sensitivity will
probably manifest itself most dramatically for nearby objects in the chemistry of prestellar disks 
around Young Stellar Objects (YSOs), and at the other end of the distance scale, 
for the chemistry of the interstellar medium of primordial
galaxies. One caveat might be mentioned: high spectral resolution is often required where there is high
spatial resolution. Even with single-dish beams of 30$^{\prime\prime}$, local galaxies such as Cen
A show disk features of only 10 km~s$^{-1}$ in width. Thus, ALMA must provide correlator capability for
high spectral resolution ($\sim$ 1 km~s$^{-1}$) at full galaxy bandwidths ($\ge$ 1000 km~s$^{-1}$)
on as many baselines as possible.\\
ALMA will make images, at 1 mm wavelength, with the same (0.1$^{\prime\prime}$), or better resolution
than achieved by the Hubble Space Telescope (HST), at visible wavelengths. This capability will
enable many new imaging programs, including:
\begin{enumerate}
\item Studies of the evolution of dust and gas from molecular clouds, to circumstellar disks, to
planetary systems;
\item Studies of the enigmatic hot cores across the Galaxy;
\item Studies of YSOs across the Galaxy;
\item Investigations of the chemistry of circumstellar disks, using multiple lines of many species,
possibly by means of line-surveys;
\item Measurements of cloud temperatures across the Galaxy, for studies of depletion in cold cores
and chemical fractionation.
\end{enumerate}

SOFIA, the new NASA airborne observatory, will give an improved view of much of the region 
of the spectrum blocked by the
atmosphere. With a 2.5 m telescope and much improved spectrometers compared to the KAO, many
specific new spectroscopic features can be searched for, with the exception of H$_2$O and O$_2$
lines, and features still hidden by the strong atmospheric lines. 
Ultimately the spectroscopy of the interstellar medium and nearby galaxies will be best studied from HIFI, 
the heterodyne instrument on Herschel, the ESA cornerstone mission. At the L2 point, an
ideal satellite location, studies such as line-surveys can be carried out continuously in time and
without interruption from H$_2$O atmospheric lines. Both SOFIA and Herschel will extend the high
quality spectroscopy range to about 2 THz or possibly more. \\

Actually, in spite of the impressive size and cost of the new instruments, much of the new
capability is provided by the smallest component: the detector element. Since their inception
($\sim$ 1979), SIS devices have increased sensitivity by a factor approaching 100. Now, essentially
quantum noise limited, the route to further improvement is in more bandwidth and more sophisticated
device structure. For instance, IF bandwidths are at about 4 GHz and are planned for as much as 20
GHz. For single-dish work, the single junction element is being replaced by devices detecting both
polarizations, having instantaneous on-off subtraction, sideband separation, and possibly balanced
mixing. Each of these features implies the use of two junctions, thus up to 16 detectors could be used for
spectroscopy of a distant point source, resulting in a sensitivity gain of about 5 or
so. Similarly, the increased frequency range ($\sim$ 2 THz) is being supported by new beam-lead, 
broad band, high output power, diode multipliers for the local oscillators.

\section{New Spectral Opportunities}

While ALMA provides sensitivity and angular resolution breakthroughs, SOFIA and Herschel (HIFI) 
generate the opportunity
for observing in new spectral ranges, due to the avoidance of some or all of the atmospheric H$_2$O. 
HIFI will provide a novel view of much of
the submillimeter spectrum, totally unperturbed by atmospheric H$_2$O. It was intended that HIFI would have total
frequency coverage from about 480 GHz to 2.7 THz, but the technical difficulty of production of the
local oscillators, combined with the limited budget and fixed launch schedule, has
forced the abandonment of the highest frequency channel, so the highest frequency now is 1.9 THz (the 
frequency of the C$^+$ line).

\subsection{Line-Surveys}

\begin{figure}[]
\resizebox{\hsize}{!}{\includegraphics{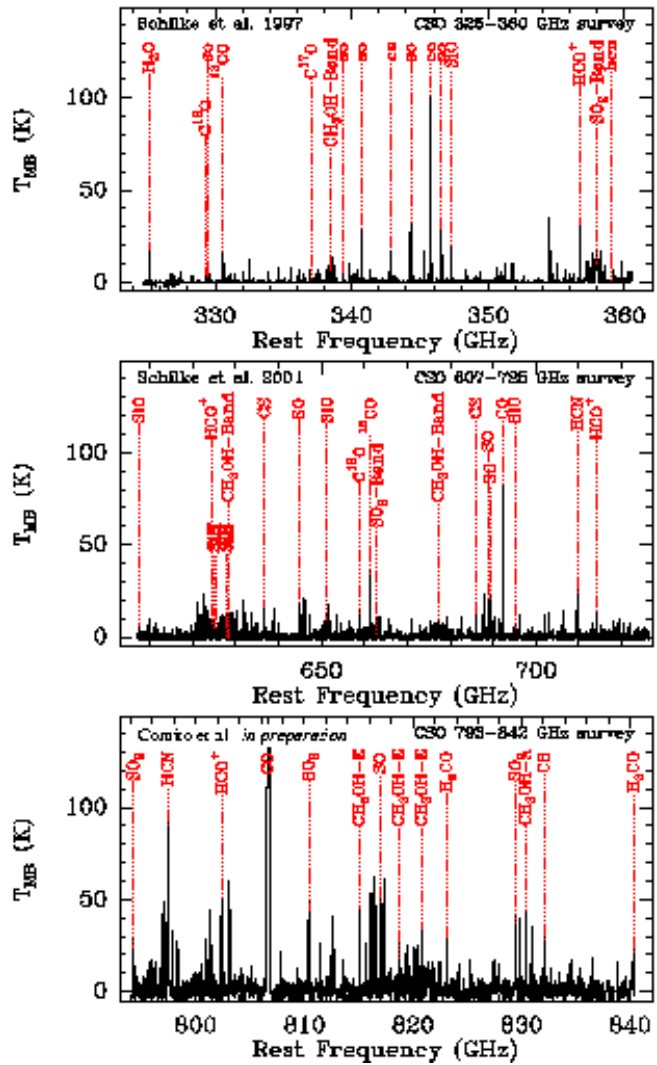}}
\caption{Line-surveys of Orion with the CSO telescope.
\label{line_survey}}
\end{figure}

Much of the HIFI data will be taken in the form of line-surveys. These have been spectacularly
successful on \\
ground-based telescopes (see Figure \ref{line_survey}), and the methodology has been
refined by means of simulations (\cite{comito01}), to make it suitable for rapid frequency
scanning, such that a full spectral survey of THz bandwidth can be completed, by HIFI, for strong sources in
a day or less. This compares with the many weeks taken on ground-based telescopes to obtain only
100 GHz of data. The factor $\sim$ 100 of improvement in time is mostly due to the automation necessary for
space and the vastly improved designs of mixers and local oscillators.

\subsection{H$_2$O}

\begin{figure}[]
\resizebox{\hsize}{!}{\includegraphics{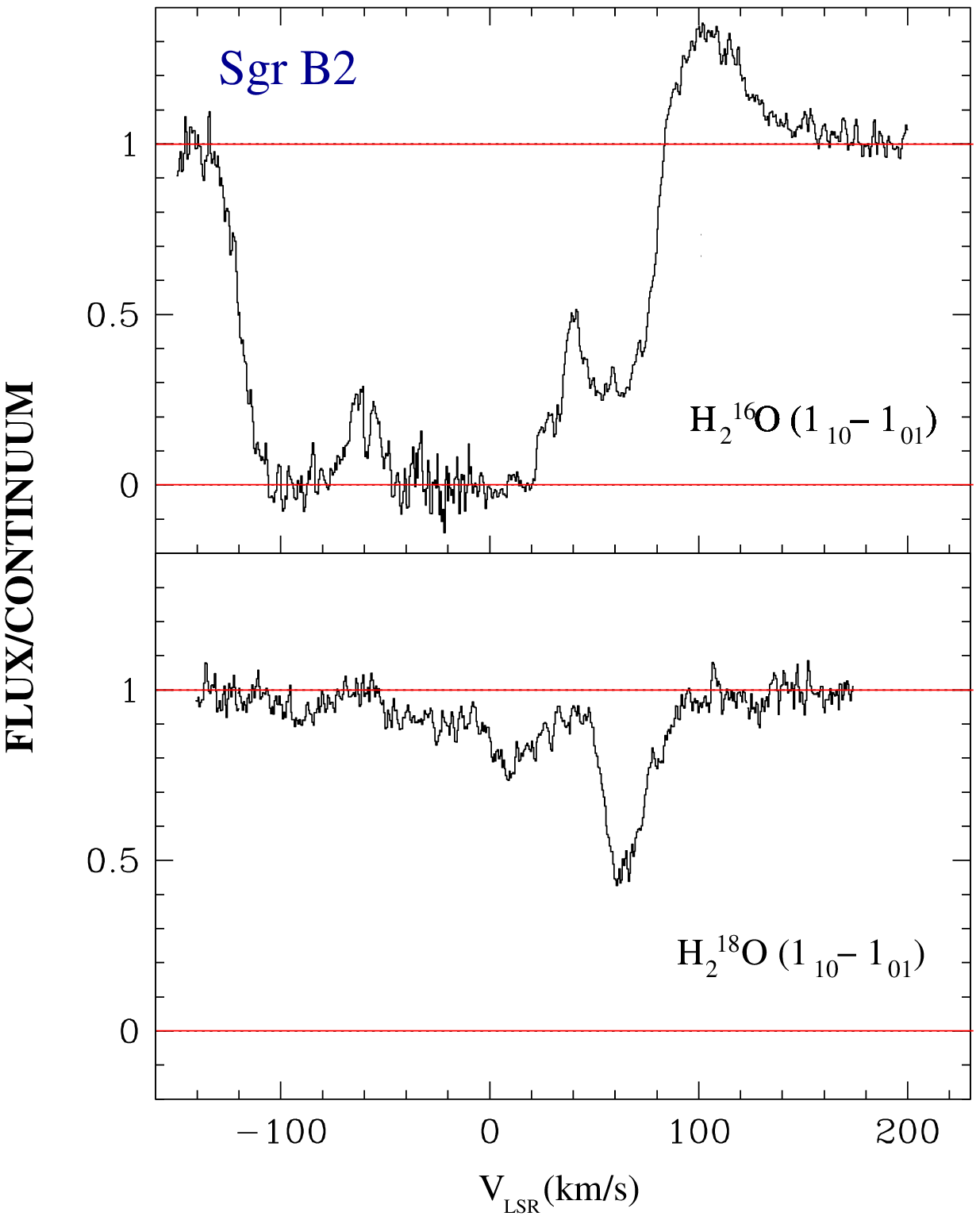}}
\caption{H$_2^{16}$O and H$_2^{18}$O line detections by the SWAS satellite toward Sagittarius B2.
\label{sgrb2}}
\end{figure}

\begin{figure}[]
\resizebox{\hsize}{!}{\includegraphics{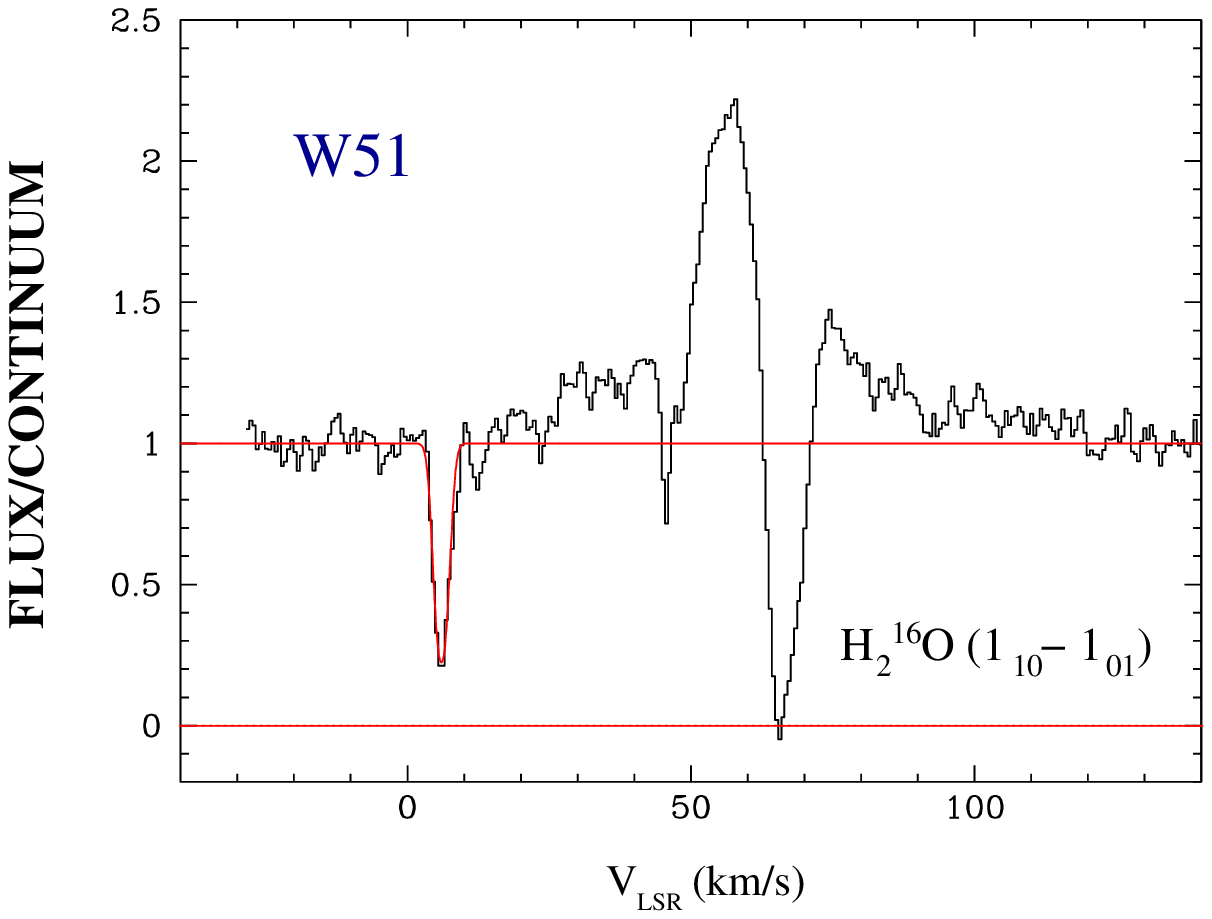}}
\caption{Two spectacular examples with W49N, a distant ($\sim$ 11.4 kpc) compact HII region whose line of 
sight crosses twice the Sagittarius spiral arm, and W51 located at $\sim$ 7 kpc.
\label{w49_w51}}
\end{figure}

Some H$_2$O lines have been observed from the ground and from the KAO, but the bulk of observations
is of the 557 GHz line from SWAS and also ODIN. Figures \ref{sgrb2} and \ref{w49_w51} 
show some amazing SWAS H$_2$O
spectra, indicating a complex structure to the line shape, often including self absorption. The three 
sources presented in these figures are Sgr B2 (\cite{neu00}), W51 (\cite{neu02}) and W49N (\cite{plume}). 
These are distant compact HII regions, where absorption lines observations can trace the water vapor content 
in clouds along their line of sight.

\begin{figure}[t]
\resizebox{9.3cm}{!}{\includegraphics{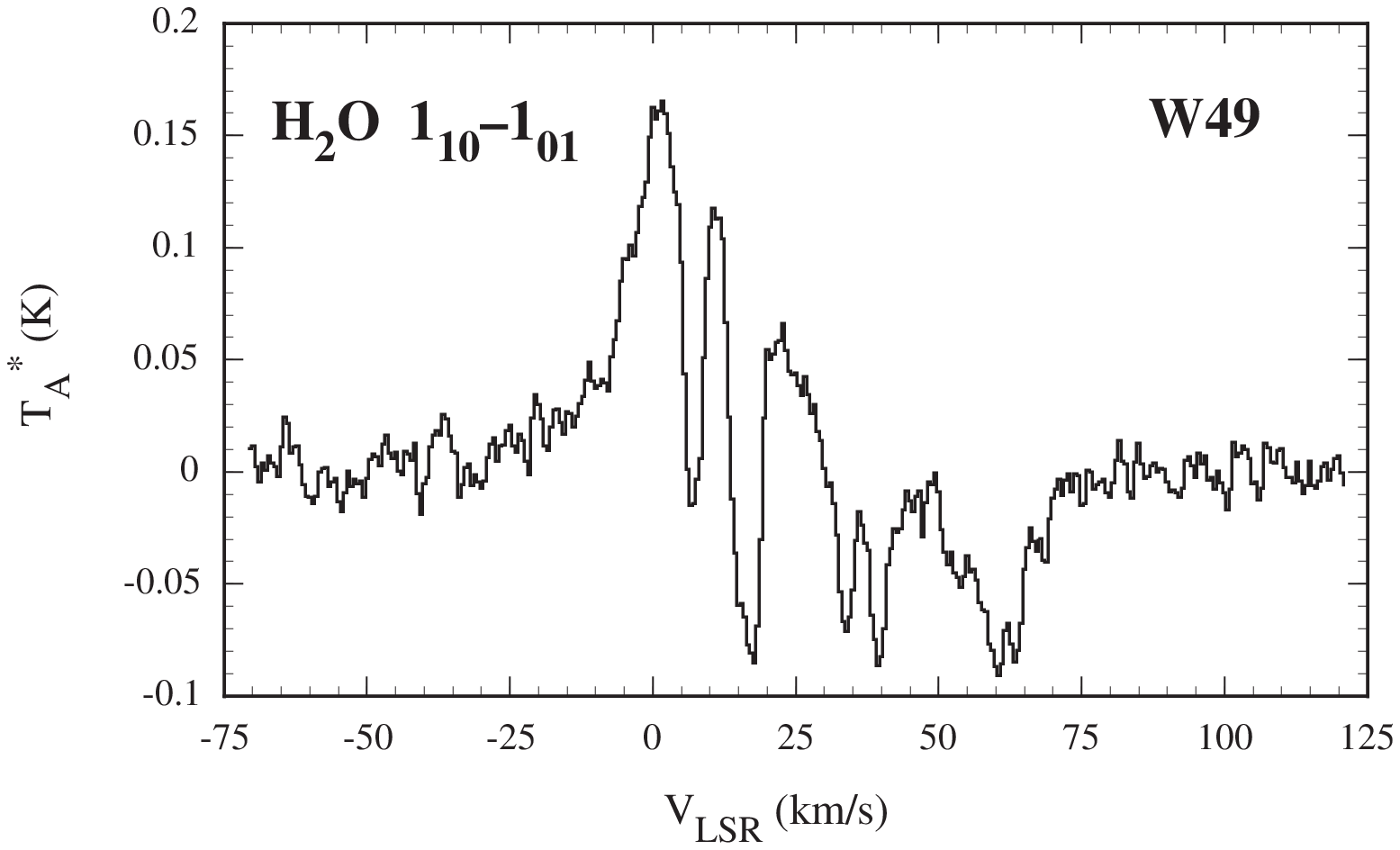}}
\end{figure}

\begin{figure*}[]
\centering
\includegraphics[width=0.8\linewidth]{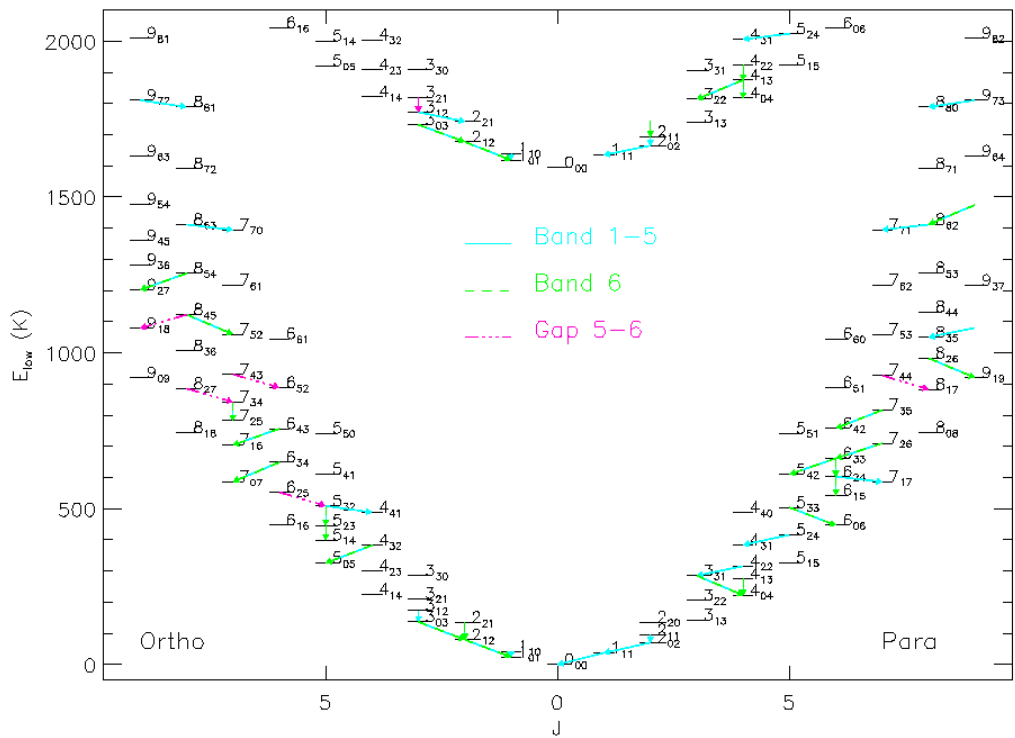}
\caption{Water transitions available to HIFI. Bands 1-5 covers the frequency range 472-1280 Gz; Band 6 covers 
1400-1904 GHz; Gap 5-6 is the frequency range between Band 5 and Band 6, which may not be covered.
\label{h2otransitions}}
\end{figure*}

HIFI will have available many transitions of H$_2$O and H$_2$$^{18}$O, most of which have not been
observed previously, including those in the important excitation path to the ground-state 
(Figure \ref{h2otransitions}).
  
These many available transitions will probe the gas cooling properties of H$_2$O in objects with a
wide range of densities, such as collapsing protostellar envelopes (\cite{ceccarelli96}), 
where H$_2$O becomes the primary coolant as the density increases in the inner regions.

\subsection{Hidden Species}

Besides the H$_2$O lines themselves, there exists a body of spectral features never observed
because of their proximity in frequency to the H$_2$O and O$_2$ lines. Some examples are lines of
H$_2$D$^+$ (e.g. 2$_{1,1}$~$\rightarrow$~2$_{1,2}$ at 1.112 THz), CH$^+$ (e.g. 1~$\rightarrow$~0 at 836 GHz), NH
(e.g. N~=~1~$\rightarrow$~0, J~=~0~$\rightarrow$~1 at 946 GHz) and LiH (e.g. 1~$\rightarrow$~0 at 444 GHz). 
Molecular lines such as HCl (1~$\rightarrow$~0) at 626 GHz and SiH at 627 GHz are hard to observe from the 
ground, being in the wings of H$_2$O lines. Many of the ``hard to detect from the ground'' species
are light, fast rotators, typified by the diatomic hydrides. Some of these, e.g. NaH, CaH,
MgH,... have been searched for in J~=~0~$\rightarrow$~1 transitions using atmospheric windows for 
ground-based telescopes, with negative
results, but can be searched for now, with increased sensitivity, in two ways. The first method (can
be carried out from the ground in favorable cases) makes use of cold clouds on the line of sight
to powerful continuum sources, such as Sgr B2. There, in the cold dark clouds, the molecules will be in
the ground state and would be detected in absorption. This method avoids the confusion in
the spectra of GMCs, since the line forest is largely made up of highly excited molecular lines,
which, of course, is not the case for the hydrides in cold clouds. A good example is seen in the SWAS 
detection of the ground-state H$_2$O line in Sgr B2 (Figure \ref{sgrb2}). The second method does suffer
from GMC line confusion, but makes use of the new high frequency capabilities for HIFI and SOFIA, where 
the high J lines of the various hydrides will have much higher line strengths than 1~$\rightarrow$~0;
hopefully this will overcome the line-forest effect in the high temperature and high density sources.\\
Generally, above 1 THz the almost zero atmospheric transmission has prevented studies so far, but several 
windows of limited size and strength have been recently detected on high sites. 
Space platforms will allow studies of HD (1~$\rightarrow$~0) at 2.67 THz, HeH$^+$ (1~$\rightarrow$~0) at 
2.01 THz, N$^+$ at 1.46 and 2.46 THz, HF at 1.23 THz, FeH at 1.41 THz, OH at 1.83 THz, OH$^+$ at 2.46 THz, 
etc... Some species have been detected by ISO , e.g. CH$^+$ in the high J transitions (\cite{cernich97}), 
HF (\cite{neu97}) and HD (\cite{wright99}, \cite{caux02}). HIFI will be a great improvement compared to what 
used to be available on ISO. 

\section{Deuterium}

One of the most important aspects of molecular spectroscopy of the interstellar medium is the study 
of the abundance of deuterium. The new facilities will provide a much wider range of frequencies and 
sensitivities, so that the deuterium studies can be extended to new species and new objects. The 
fundamental aspect of deuterium chemistry is that the chemical fractionation process (see below) 
forces deuterium atoms into heavy molecules, at the expense of hydrogen, whenever the medium is 
cold, with amazing results.\\
Deuterium-bearing molecules have become the target of many observations in
recent years and several models have been developed to account for
them (\cite{tielens83}, \cite{roberts00a}, \cite{roberts00b}). 
More than twenty such molecules have been detected to date in
interstellar clouds with abundances, relative to the non-deuterated
counterpart, ranging from 10$^{-1}$ to 10$^{-4}$. It is accepted that
deuterium is produced during the Big Bang and it is generally believed
that since the Big Bang, deuterium has been destroyed but not created
in nuclear reactions occurring inside stars. \\
The most reliable determinations of the D/H ratio are based on
spectroscopic measurements of Lyman series ultraviolet absorption
lines from foreground interstellar gas. In our Galaxy, this has been
obtained (via DI and HI observations) with satellites such as the
International Ultraviolet Explorer (IUE), Copernicus, the Extreme
UltraViolet Explorer (EUVE), the Hubble Space Telescope (HST) and
recently the Far Ultraviolet Spectroscopic Explorer (FUSE). 
Measurements of the D/H ratio toward high-redshift systems like
quasars (e.g. \cite{tytler99}) seemed to show more dispersion than expected and an inverse 
correlation of this abundance with HI column
density. If this is real, it would suggest that in these high HI
column density systems, some processing of D/H must have occurred (\cite{vidal}, \cite{fields01}).
It is interesting to note that abundances of deuterium measured in the
interstellar medium also appear to show considerable dispersion. From
published values, D/H ranges from $\sim$ 5 $\times$ 10$^{-6}$ to
$\sim$ 4 $\times$ 10$^{-5}$.
FUSE observations of seven white dwarfs and sub-dwarfs lead to a D/H ratio of
(1.52 $\pm$ 0.08) $\times$ 10$^{-5}$ (\cite{moos02}), to be compared
with the value of (1.5 $\pm$ 0.1) $\times$ 10$^{-5}$ (\cite{lin98})
determined from HST observations of late-type stars. Both measurements
refer to warm interstellar gas, located within 100 pc of the
Sun. However, differences by a factor of two have been derived from
Copernicus and HST data toward stars located between 100 and 500 pc
from the Sun. The ratio seems constant within 100 pc, but seems to
vary at considerably greater distances.\\
There is no identified process which can explain such large variability 
and without an understanding it is not justified 
to use an average D/H ratio to represent the primordial deuterium abundance. 
We will argue, below, that the chemistry of the interstellar medium could be 
responsible, in that it can extract large amounts of deuterium which becomes trapped 
in molecules and on grains. 

\subsection{Basic Chemistry}

Deuterium bearing species are good probes of the cold phases of
molecular clouds prior to star formation and many recent observations
point to the fact that their abundance relative to their
hydrogenated analogues are larger, by a factor up to 10000, than the
solar neighborhood value of $\sim$ 1.5 $\times$ 10$^{-5}$ (see references above). \\
Therefore the relative abundance of isotopomers does not measure the relative 
abundances of the isotopes themselves. The chemical fractionation process arises
from differences in the molecular binding energies caused by the different zero-point 
vibration energy. Almost incredibly, this can lead to a detectable quantity of the triply 
deuterated ammonia (see section 4.2).\\

\begin{figure}[h]
\resizebox{\hsize}{!}{\includegraphics{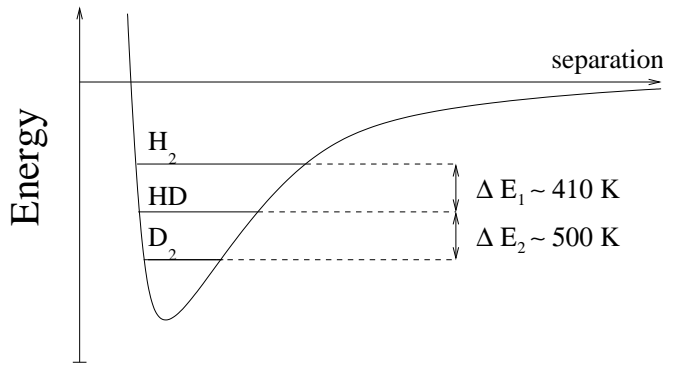}}
\caption{H$_2$, HD and D$_2$ potential energy diagram. $\Delta$E$_i$ is the
difference between the zero point energies relative to the minimum of
the molecular potential curve.
\label{nrj_well2}}
\end{figure}

In molecular clouds, hydrogen and deuterium are predominantly in the
form of H$_2$ and HD respectively. So the HD/H$_2$ ratio should closely
equal the D/H ratio. Since the zero-point energies of HD and H$_2$
differ greatly (see Figure \ref{nrj_well2}), the chemical
fractionation will favor the production of HD compared to H$_2$.\\

Deuterium is initially removed from the atomic phase through charge exchange with H$^+$, 
followed by reaction with the abundant H$_2$. HD could further interact with D$^+$ again to 
give D$_2$:
\begin{equation}
H^+ + D  \longleftrightarrow H + D^+  
\end{equation}
\begin{equation}
D^+ + H_2  \longleftrightarrow HD + H^+ + \Delta E_1  
\end{equation}
\begin{equation}
D^+ + HD  \longleftrightarrow D_2 + H^+ + \Delta E_2  
\end{equation}

The reactions 2 and 3 are exothermic as substituting an 
H atom versus an D atom in a polyatomic 
molecule generally leads to a gain in energy. These energies may be computed (at 0 K) 
by the differences between the zero-point energies of the products and the reactants. The energies 
$\Delta$E$_1$ and $\Delta$E$_2$ are quoted in figure \ref{nrj_well2}.
   
In the dense, cold regions of the 
interstellar medium, D will be initially nearly all absorbed into HD. The abundant ion available 
for interaction is H$_3^+$, which gives H$_2$D$^+$:
\begin{equation}
H_3^+ + HD  \longleftrightarrow H_2D^+ + H_2 + \Delta E_3 
\end{equation}
where $\Delta$E$_3$/k $\sim$ 230 K for a typical temperature of a dark cloud 
of about 10 K (e.g. \cite{millar89}). The reverse
reaction does not occur efficiently in the cold dense clouds where 
obviously the temperature is much lower than $\Delta$E$_3$. Therefore, the
degree of fractionation of H$_2$D$^+$ becomes non-negligible.\\
This primary fractionation can then give rise to a second
fractionation:
\begin{equation}
H_2D^+ + CO \longleftrightarrow DCO^+ + H_2
\end{equation}
\begin{equation}
\hspace{2.05cm} \longleftrightarrow HCO^+ + HD
\end{equation}

In dark clouds H$_3^+$ gives rise to HCO$^+$ via the reaction:
\begin{equation}
H_3^+ + CO \longleftrightarrow HCO^+ + H_2
\end{equation}

\subsection{Multiply Deuterated Molecules}

The study of doubly deuterated interstellar molecules has been booming in the last few 
years since the surprising discovery of a large amount ($\sim$ 5\%) of doubly deuterated 
formaldehyde in the low mass protostar IRAS 16293-2422 (\cite{ceccarelli98}, 
\cite{loinard00}). This is more than one order of magnitude higher than in Orion KL 
where D$_2$CO was first detected by \cite*{turner90}. This first discovery was 
followed by many other studies which confirmed the presence of very large amount of 
doubly deuterated formaldehyde (D$_2$CO) as well as ammonia (ND$_2$H) (e.g. \cite{roueff00}, 
\cite{loinard01}). 
Gas phase chemical models account relatively well for the observations of
the abundances of singly deuterated molecules but may not be able to completely 
reproduce the large deuterations observed for multiply deuterated
molecules. The large deuterations could also be a
product of active chemistry on the grain surfaces as predicted by
\cite{tielens83} with two processes:
\begin{enumerate}
\item Deuteration during the mantle formation phase;
\item Evaporation of the mantles ices resulting from the heating of
the newly formed star, with injection into the gas phase of the
deuterated species.
\end{enumerate} 

\begin{figure}[h]
\resizebox{\hsize}{!}{\includegraphics[]{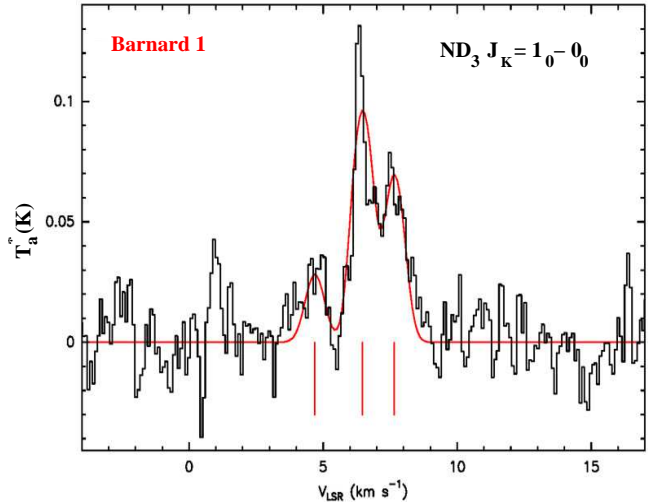}}
\caption{Spectrum of the 309.91 GHz ND$_3$ line towards Barnard 1 observed at the CSO. The red lines show the 
fits to the position and strength of the hyperfine components.
\label{nd3}}
\end{figure}

\begin{figure}[h]
\resizebox{\hsize}{!}{\includegraphics[angle=270]{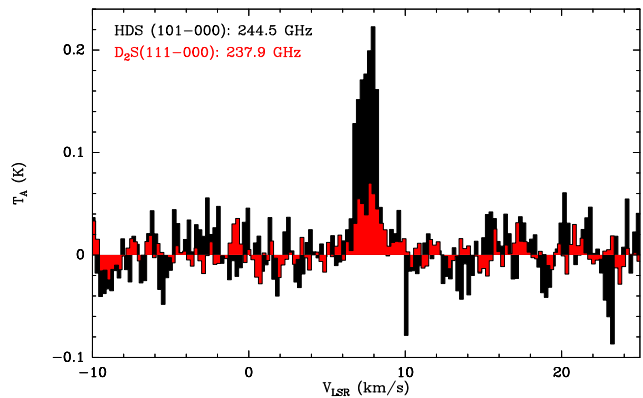}}
\caption{HDS (1$_{0,1}$~$\rightarrow$~0$_{0,0}$) and D$_2$S (1$_{1,1}$~$\rightarrow$~0$_{0,0}$) detections with the CSO 
towards the DCO$^+$ (3~$\rightarrow$~2) emission peak of NGC 1333 IRAS 4A (Vastel et al., {\it in preparation}). 
\label{iras4a}}
\end{figure}

The year 2002 appears to be the cornerstone in the study of deuteration processes with the 
first detection of a 
triply deuterated molecule, ND$_3$. Until now, the possibility for detecting triply deuterated 
molecules was so remote that their lines were omitted in the spectroscopic catalogs for 
astrophysics. The ground-state rotational transition at 309.91 GHz of ND$_3$ has been 
detected with the CSO towards the Barnard 1 cloud (\cite{lis02}, see Figure \ref{nd3}) 
and the NGC 1333 IRAS 4A region (\cite{vdt02}), with 
abundance ratios [ND$_3$]/[NH$_3$] $\sim$ 10$^{-3}$ and [ND$_3$/H$_2$] $\sim$ 10$^{-11}$. 
They conclude that reactions in the gas-phase are more likely to produce these high degrees of 
deuteration, rather than grain surface chemistry. However they cannot totally rule out the possibility 
that surface processes also contribute to the formation of ND$_3$ (cf \cite{rodgers01}). 
Moreover, the recent discovery of doubly-deuterated methanol towards IRAS 16293-2422 
(CHD$_2$OH/CH$_3$OH $\sim$  0.2, \cite{parise}) cannot be accounted for gas-phase models. 
The most promising route for such methanol deuteration seems to be by surface chemistry.
Recent observations of the extended D$_2$CO emission towards L1689N (\cite{ceccarelli02}) pointed 
out the difficulty in explaining the D$_2$CO abundance, either with gas phase models or with grain surface 
chemistry. 
The many routes to deuteration are not fully understood and more observations are necessary to 
elucidate theses processes. 
Very recently, doubly deuterated hydrogen sulfide has been detected by \cite*{vastel} with the CSO towards 
the Barnard 1 cloud and the DCO$^+$ (3~$\rightarrow$~2) emission peak of NGC 1333 IRAS 4A. 
Figure \ref{iras4a} shows as an 
example the first detection of D$_2$S (1$_{1,1}$~$\rightarrow$~0$_{0,0}$) together with HDS 
(1$_{0,1}$~$\rightarrow$~0$_{0,0}$).

\subsection{Depletion of CO}

Gas phase species are expected to be depleted at the centers of cold,
dark clouds, since they tend to stick to the dust grains. 
A series of recent observations has shown that, in some cases, the
abundance of molecules like CO decreases toward the core center of
cold, dense clouds (L1498: \cite{willacy98}; IC 5146: \cite{kramer99}, \cite{bergin01}; 
L977: \cite{alves99}; L1544: \cite{caselli99}; L1689B: \cite{jessop01}; \cite{bacmann02}).  
These decreases in abundance have been interpreted as resulting from the depletion of molecules
onto dust grains (see, e.g., \cite{bergin97}; \cite{charnley97}). It is
now clear that these drops in abundance are typical of all dense cores.
The removal of these reactive species affects the gas-phase chemistry and
particularly the deuterium fractionation within the cloud. Indeed, the
removal of species which would normally destroy H$_3^+$ (e.g. CO; \cite{roberts00a}) means
that the H$_3^+$ is more likely to react with HD and produce
H$_2$D$^+$ (reaction 4). \\

H$_2$D$^+$ has been detected toward the low luminosity
proto-stars NGC 1333 IRAS 4A (\cite{stark99}) and IRAS 16293-2422
(\cite{stark}) with a H$_2$D$^+$/H$_2$ ratio of 3 $\times$ 10$^{-12}$ 
in the former. Recently, \cite*{caselli} detected the H$_2$D$^+$ 
372 GHz line towards two pre-stellar cores, L1544 and L1521F, both
probably on the verge of collapse. 

\subsection{Modification of the Deuterium Chemistry}

\begin{figure}[h]
\resizebox{\hsize}{!}{\includegraphics{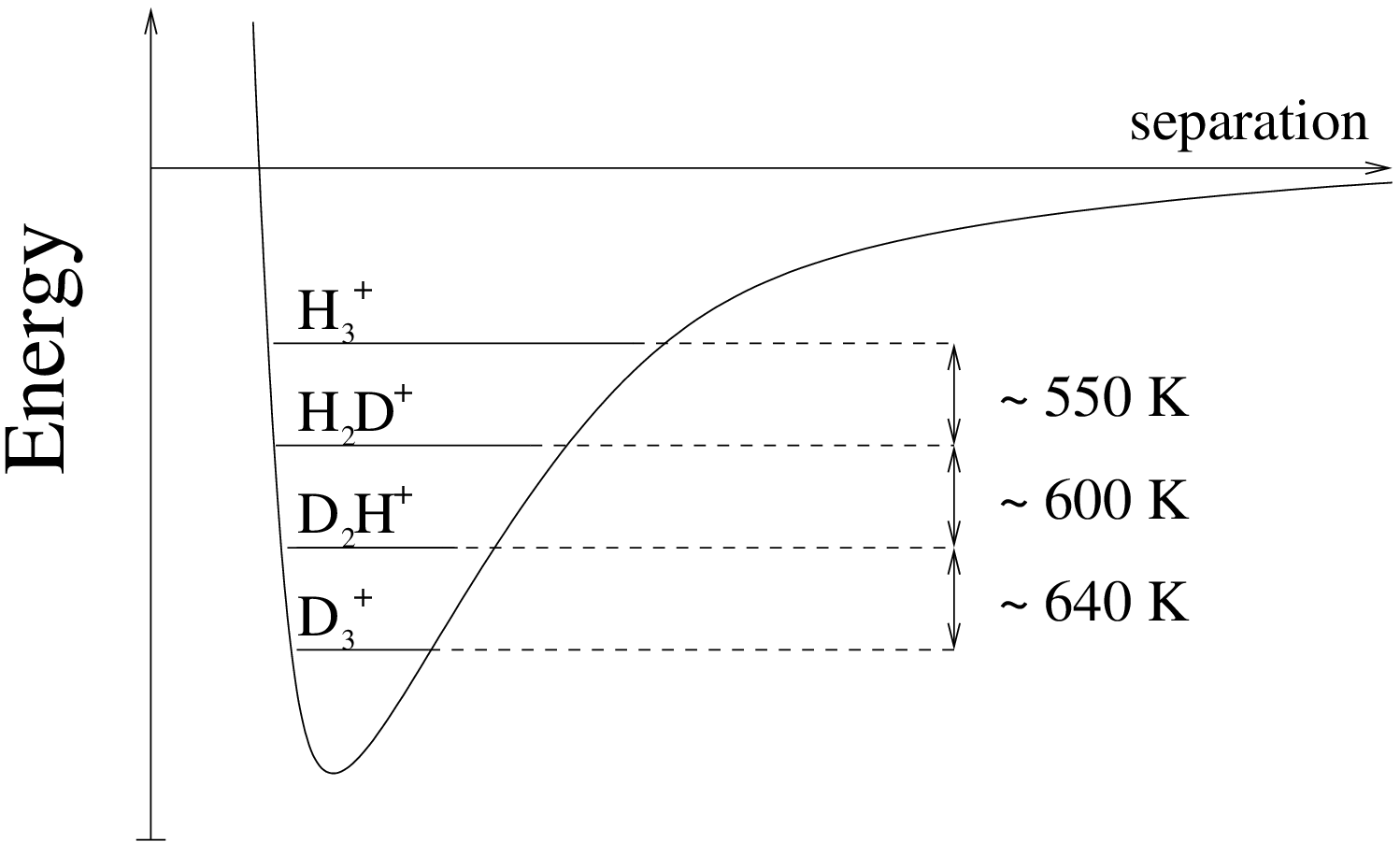}}
\caption{H$_3^+$, H$_2$D$^+$, D$_2$H$^+$ and D$_3^+$ potential energy diagram. 
\label{nrj_well3}}
\end{figure}

The large differences found in the values of the deuterium abundance in the interstellar medium pose
the question as to the mechanisms responsible for these variations. A possible answer is the
chemical fractionation process, which, in cold regions of the interstellar medium, steadily forces
the deuterium into the heavy molecules. The trend, is to minimize the free energy, which implies 
forming the heaviest species, with the least uncertainty energy (see Figure \ref{nrj_well3}).
As shown in the previous section, in the very cold ($\sim$ 10 K) regions where, e.g. ND$_3$ is 
detected, CO is in fact heavily depleted by accretion on to grains. For example, 
[CO/H$_2$]~$\sim$~5 $\times$ 10$^{-6}$ (\cite{bacmann02}), leaving HD at 
[HD/H$_2$]~$\sim$~5 $\times$ 10$^{-5}$ as the most abundant molecule available for reaction with H$_3^+$ and 
H$_2$D$^+$. Another example is the case of HD 112 $\mu$m absorption towards a cold molecular cloud in 
the line of sight of W49 where \cite*{caux02} found that HD is 12 times more abundant than CO. 
The result may be the production of more high deuterium content molecules: 
\begin{equation}
H_3^+ + HD \longleftrightarrow H_2D^+ + H_2 + \Delta E_a
\end{equation}
\begin{equation}
H_2D^+ + HD \longleftrightarrow D_2H^+ + H_2 + \Delta E_b
\end{equation}
\begin{equation}
D_2H^+ + HD \longleftrightarrow D_3^+ + H_2 + \Delta E_c
\end{equation}

where $\Delta$E$_a$, $\Delta$E$_b$ and $\Delta$E$_c$ are the released energies of the exothermic 
reactions. Using the zero-point energies computed by \cite*{carney80}, and the energy of the 
first allowed rotational state of the H$_3^+$ molecule permitted by the Pauli exclusion principle 
($\sim$ 92 K), these values are: $\Delta$E$_a$ = $\sim$ 230 K, $\Delta$E$_b$ = $\sim$ 180 K and 
$\Delta$E$_c$ = $\sim$ 230 K.  

The species D$_2$H$^+$ may indeed be detectable in sufficiently cold, CO depleted regions. Because of 
the low temperature of the highly deuterated regions ($\sim$ 10 K), the lowest lying transition of 
H$_2$D$^+$ and D$_2$H$^+$ will be most appropriate and can be searched for in absorption against 
Sagittarius B2, possibly:
\begin{itemize}
\item H$_2$D$^+$$_{1_{01}-0_{00}}$ at 1370.15 GHz $^*$
\item H$_2$D$^+$$_{2_{11}-2_{12}}$ at 1111.74 GHz
\item D$_2$H$^+$$_{1_{11}-0_{00}}$ at 1476.60 GHz
\item D$_2$H$^+$$_{2_{20}-2_{11}}$ at 1370.05 GHz $^*$
\end{itemize}

$^*$ It appears to be purely chance that these frequencies are so close.

\subsection{D, HD Depleted by Fractionation?}

It would be a difficult task to make an inventory of all D substituted molecular species throughout 
the Galaxy, to see if the measured values of the deuterium abundance could be affected. The 
fractionation might work in favor of metal species (e.g. ND$_3$) in high metallicity galaxies, or 
in favor of purely hydrogenic species (e.g. H$_2$D$^+$ and D$_2$H$^+$) in low metallicity objects. 
At any rate, we can ask if it is physically possible for the former case to occur. If we take the local 
value of C, N and O abundances and count the total number of bonds available for substitution of H by D, 
we get 3~$\times$~10$^{-3}$~[H]. But [D]/[H] is only $\sim$~10$^{-5}$, so there does exist the possibility 
of such an effect. In fact, if the product of the fraction of available bonds actually occupied by D 
(f$_D$) times the fraction of the interstellar medium which is cold enough for strong fractionation and 
CO depletion (f$_C$) approaches 10$^{-3}$, then there will be a significant effect:
\begin{equation}
f_D \times f_C \ge 10^{-3}
\end{equation}
We know that f$_D$~$\sim$~10$^{-1}$, or more in some cases, so if f$_C$~$\ge$~10$^{-2}$, measured values 
of the deuterium abundance may be significantly affected.

  
\begin{acknowledgements}

We acknowledge support from the NSF under contract AST-9980846. We thank Eric Herbst, John Pearson, 
Tom Millar and Dieter Gerlich for helpful conversations. We are also grateful to Paul 
Goldsmith for providing us the figures of SWAS observations towards Sgr B2, W49N and W51.

\end{acknowledgements}

\end{document}